\documentclass[12pt,a4paper]{article}

\usepackage[utf8]{inputenc}
\usepackage{textcomp}

\usepackage{amsfonts}
\usepackage{amssymb}

\usepackage{mathrsfs}
\usepackage{amsmath}
\usepackage{amsfonts}
\usepackage{amsthm}
\usepackage{mathabx}

\usepackage{graphicx}
\usepackage{subfigure}

\begin{document}
\title{Constraints on ADM tetrad gravity parameter space from S2 star in the center of the Galaxy and from the Solar System}
\author{Mattia Villani \\ INFN sez. Firenze}
\date{}
\maketitle
\abstract{ADM tetrad gravity is an Hamiltonian reformulation of General Relativity which gives new insight to the Dark Matter Problem. We impose constraints on the parameter space of ADM tetrad gravity with a Yukawa-like ansatz for the trace of the extrinsic curvature of the 3D hypersurfaces by fitting the orbit of the  S2 star around the Black Hole in the Galactic center and using the perihelia of some of the planets of the Solar System. We find very thight constraints on the \emph{strength} of the coupling, $4.2 \,\times \,  10^{-4} \, \text{AU}\,\lesssim \, \delta \, \lesssim \, 4.6 \, \times \, 10^{-4} \, \text{AU}$, and an upper limit for the (inverse) scale length, $\mu \, \lesssim \, 3.5 \, \times \, 10^{-6} \, \text{AU}^{-1}$.}

\section{Introduction}
ADM tetrad gravity (ADM tg) \cite{adm1, adm2,adm3,adm4,adm5,adm6,adm7} (first introduced in \cite{adm8}) is an Hamiltonian reformulation of General Relativity based on a 3+1 splitting of spacetime into 3-dimensional spacelike hypersurfaces parametrized by the time,\footnote{These hypersurfaces are simultaneity surfaces.} just like in the ADM formulation \cite{adm}, but it considers tetrads as the dynamical variables instead of the metric. 

In \cite{adm6,adm7}, authors studied Post Minkowskian (PM) and Post Newtonian (PN) expansion, in particular they calculated the equation of motion for a particle of mass $m$ subjected to the gravitational force due to a potential $\Phi$, showing that there is a  0.5 PN (i.e. at the order $O(c^{-1})$) correction depending on the trace of the extrinsic curvature  and on the velocity of the particle:
\begin{equation}\label{eq:motion}
m \, \ddot{\vec{x}} = -  m \, \vec{\nabla} \, \Phi  - m \,\dfrac{\dot{\vec{x}}}{c} \, \dfrac{d^2}{dt^2} \, {}^3\widetilde{\mathcal{K}}_{(1)}(c \, t, \vec{x})
\end{equation}
where:
\begin{equation}
{}^3\widetilde{\mathcal{K}}_{(1)} = \int{\dfrac{{}^3K_{(1)}(ct,{\vec{x}}^\prime)}{|{\vec{x}}^\prime - \vec{x}|} \, d^3x^\prime}
\end{equation}
with ${}^3K_{(1)}$ being the first order trace of the extrinsic curvature. 

Rearranging equation \eqref{eq:motion}, we have:
\begin{equation}\label{eq:motion2}
\dfrac{d}{dt} \, \left[ m \,  \left( 1 + \dfrac{1}{c} \, \dfrac{d}{dt} \, {}^3\widetilde{\mathcal{K}}_{(1)}(c \, t, \vec{x}) \right) \, \dot{\vec{x}} \right]= - m \, \vec{\nabla} \, \Phi
\end{equation}
showing that the 0.5 PN correction can be interpreted also as an additional (position and velocity dependent) mass, thus giving an interesting and new insight on the Dark Matter problem, which, in this formalism, is an inertial effect due to the shape of the simultaneity hypersurfaces.\footnote{As has been pointed out in \cite{adm6,adm7}, in this way, the Dark Matter becomes  a metrology problem, since the trace ${}^3K$ and the shape of the hypersurfaces are linked to the choice of the convention on clock synchronizations.}\footnote{We note further, that equation \eqref{eq:motion2} implies a violation of the Equivalence principle in the 3-space (\emph{not} in the four dimensional spacetime).}

\vspace{2mm}
As has been proved in \cite{adm4}, in the ADM tg formalism, ${}^3K$ is a gauge variable, but there is no known \emph{natural} gauge  for this function: in this paper, we first discuss our ansatz for ${}^3\widetilde{\mathcal{K}}_{(1)}$. 

In section \ref{sec:fit}, we fit equation \eqref{eq:motion} to the data of the star S2 orbiting the black hole in the center of the Galaxy described in \cite{gill1,gill2}; we will then find the part of the parameter space in which ADM tg gives a better fit than Newtonian gravity: this will impose limits on the free parameters of the theory (this is similar to what \cite{capo} did with $f(R,\phi)$ theories with a Sanders-like potential, \cite{sand}). 

In order to improve our constraints, in section \ref{sec:sol_sys}, we derive a formula for the precession angle given by the 0.5 PN correction and use it and perihelia precession angles of some of the planets of the Solar System to impose upper limits to our free parameters. 

Finally, in section \ref{sec:both}, we combine those constraints.

\section{Our ansatz for ${}^3\widetilde{\mathcal{K}}_{(1)}$}
As we said in the introduction, in \cite{adm4}, it was shown that in ADM tg the trace of the extrinsic curvature is a gauge variable, but no natural gauge is known; we shall now suggest a possible ansatz, at least for the first order ${}^3K_{(1)}$.

In \cite{adm6,adm7}, authors also calculate the PN expansion of the metric, showing that the time-time component is given by ($\Phi$ is the Newtonian potential):
\begin{equation}\label{eq:met}
{}^4g_{tt} = 1 - 2 \,\dfrac{\Phi}{c^2} \, - \dfrac{2}{c} \, \dfrac{\partial}{\partial \, t }\, {}^3\widetilde{\mathcal{K}}_{(1)}(c \, t, \vec{x}).
\end{equation}

It is known (see the review \cite{capo2} and references therein), that the PN expansion of $f(R)$ theories has the form:
\begin{equation*}
{}^4g_{tt} = 1 - 2 \, \dfrac{GM}{c^2} \, \dfrac{1}{r} \, \dfrac{1}{1+\delta} \, \Big[ 1 + \delta \,  \exp{\left( - \mu \, r \right)} \Big]
\end{equation*}
where the constant $\delta$ is the measure of the Yukawa coupling.

Confronting the previous equations one can make the ansatz that the trace of the extrinsic curvature has the form:
\begin{equation*}
{}^3\widetilde{\mathcal{K}}_{(1)} = \Delta(c \, t) \, \dfrac{1}{r} \, \exp{\left( - \mu \, r \right)}
\end{equation*}
and one can assume that the time dependence is linear $\Delta(c \, t) = c \, t \,\delta$, so our ansatz is:
\begin{equation}\label{eq:ansatz2} 
{}^3\widetilde{\mathcal{K}}_{(1)} = c \, t \, \delta\, \dfrac{1}{r} \, \exp{\left( - \mu \, r \right)}.
\end{equation}
In this way \eqref{eq:met} becomes:
\begin{equation}
{}^4g_{tt} = 1 - 2 \,\dfrac{\Phi}{c^2} \, - 2 \, \delta\; {}^3\widetilde{\mathcal{K}}_{(1)}(c \, t, \vec{x}).
\end{equation}

There are two free parameters in \eqref{eq:ansatz2}: the \emph{strength} of the Yukawa coupling, $\delta$, and the (inverse) length scale, $\mu$; the aim of this work is to impose constraints on these parameters.

\section{Constraints from the S2 star}
In this section, we use publicly available data (see the complementary material of \cite{gill2}) of the orbit of the S2 star around the black hole in the center of the Galaxy to reduce the parameter space, in a similar fashion to \cite{capo}.
\label{sec:fit}
\subsection{Method}
We fit the orbit given by the equation \eqref{eq:motion} to the data reported in \cite{gill1,gill2}, using the ansatz discussed in the previous section (eqn. \eqref{eq:ansatz2}).

We use the following algorithm, written with \emph{Wolfram Mathematica} 8, to carry out our fits:
\begin{enumerate}
\item We start with an educated guess for the initial position of the star (the initial position and velocity of the star);
\item With those data we calculate the orbit of the star numerically with a $4^{th}$ order Runge Kutta, obtaining the theoretical position of the star: $x_{th}$,\dots;
\item We minimize the $\chi^2$ (see \cite{capo}): \[ \sum_i \, \left[ \left( \dfrac{x_{i} - x_{th}(i)}{dx_i} \right)^2 + \left( \dfrac{y_{i} - y_{th}(i)}{dy_i} \right)^2 \right] \]
where $x_i$ and $y_i$ are the observed positions and $dx_i$ and $dy_i$ are the relative errors; the index $i$ runs over the obsevations. The parameters used in the fit are the initial conditions of the star and the \emph{strength} of Yukawa coupling, $\delta$, and it scale length, $\mu$.
\item Errors are calculated using the Fisher Matrix method (see also \cite{gill1} and \cite{cpp}).
\end{enumerate}

In our simulations we followed \cite{gill1,gill2} and in the Newtonian potential we fixed the mass of the central black hole (supposed fixed and point like) to $M_{bh} = 4.3 \, \times \, 10^6$ M$_\Sun$; we also fixed the distance of the Sun from the center of the Galaxy to $R_0 = 8.3$ kpc. 

The results of our fits are given in the next section.

\subsection{Results}

As a result of our fits, we obtain $\chi^2 = 1.5477$ for a Keplerian orbit and $\chi^2 = 1.516890$, with two additional degrees of freedom, for ADM tg: so the latter gives only a marginally better fit to the orbit. 

\begin{figure}[htb]
\centering
\subfigure[Keplerian orbit]{\includegraphics[scale=0.69]{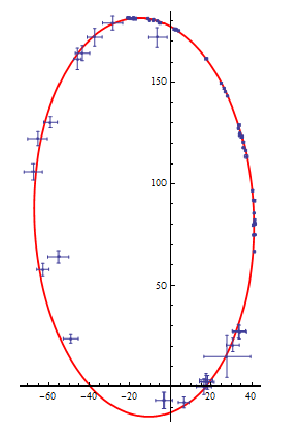}}
\subfigure[ADM tg orbit]{\includegraphics[scale=0.75]{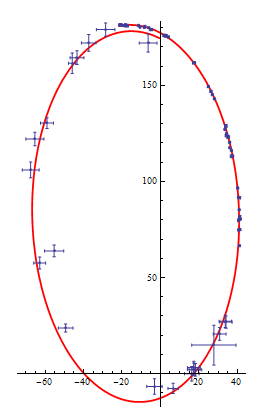}}
\caption{Fit to the orbit of the S2 star: fitted data and relative errors are in blue, the red line is the orbit.}\label{fig:orbit}
\end{figure}

Orbits are shown in figure \ref{fig:orbit}; the best fit values for $\delta$ and $\mu$ are given in table \ref{tab:res} together with their errors.

\begin{figure}[h!]
\centering
\subfigure[Light green: region of parameter space where ADM tg gives a better fit to the orbit then Newton gravity.]{\includegraphics[width=0.42\textwidth, keepaspectratio]{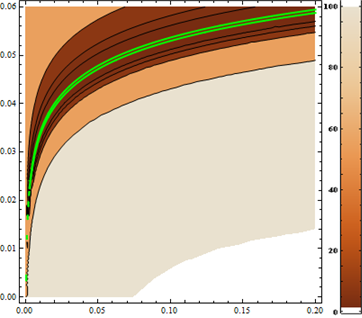}}
{$\phantom{spa}$}
\subfigure[A zoom in the region: $3 \, \times \, 10^{-4} < \delta\, \text{(AU)}  <5.5 \, \times \, 10^{-4}$ and $10^{-8}  \, < \mu  \, (\text{AU}^{-1})< 10^{-7}$ \label{fig:constr1}.]{\includegraphics[width=0.45\textwidth, keepaspectratio]{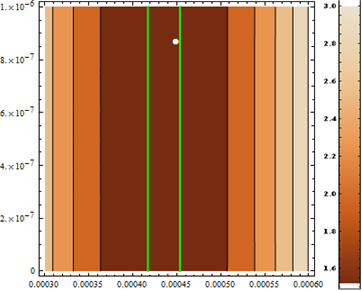}}
\caption{In both figures: on the abscissa, the $\delta$ on the ordinate $\mu$. Green line $\chi^2 = 1.5477$ (our reduced $\chi^2$ for Newtonian gravity).}\label{fig:constr}
\end{figure}

In figure \ref{fig:constr} we plot the parameter region in which ADM tg gives a better fit to the orbit than Newton's gravity: in green we plot the region where $\chi^2 = 1.5477$, the white dot is our best fit. 

We see that the parameters are not well constrained: we only have that $\delta \gtrsim 4.0 \, \times \, 10^{-4}$ AU (see figure \ref{fig:constr1}). 

In the section \ref{sec:both}, we shall combine these constraints with other coming from the precession or perihelia of some of the planets of the Solar System: this will leave available only a small region of parameter space.

\begin{table}[h!]
\centering
\begin{tabular}{c|cc}
\hline
\multicolumn{3}{c}{Keplerian orbit}\\
\hline
\multicolumn{2}{c|}{$\chi^2$} & 1.5477 \\
\hline
\multicolumn{3}{c}{ADM tg orbit}\\
\hline
Parameter & Value & Error \\
$\delta$ (AU) & 0.00045 & $0.0010$\\
$\mu$ (AU)$^{-1}$& $8.7\,\times\,10^{-7}$ & 0.04\\
\hline
\multicolumn{2}{c|}{$\chi^2$} & 1.516890 \\
\hline
\end{tabular}
\caption{Fit results: the reduced $\chi^2$ for the Keplerian orbit and for ADM tg (we remind that the ADM tg has two more degrees of freedom). For ADM tg we give also the best fit values and the respective errors for the \emph{strength} of the Yukawa coupling $\delta$ and its legth scale $\mu$. \label{tab:res}}
\end{table}

\section{Constraints from the Solar System}\label{sec:sol_sys}
In \cite{sol2,sol3,sol4} the precession of perihelia is used to set constraints on the density of Dark Matter in the Solar System: the correction to the precession coming from Dark Matter must be at most of the same order of magnitude as the error on the measurements, thus imposing un upper value to its density; here we do the same:\footnote{After all we are aiming to describe Dark Matter as a manifestation of the non Euclidicity of the hypersurface.} the correction to precession angle given by our 0.5 PN term (equation \eqref{eq:precessione}, in the next section) will imopose upper limits on the couple $(\delta, \mu)$. 

We shall now derive a formula for the precession angle given by our 0.5 PN term.

\subsection{Precession angle}
As can be seen from equations \eqref{eq:motion}, our perturbing force depends on the velocity $\dot{\vec{x}}$ of the star, so it is \emph{not a central force}. We follow  \cite{prec1} and use the \emph{Hamilton vector}, since this mehod is easily generalized to our case (see \cite{ham1,ham2,ham3,ham4,ham5} for the definition of the Hamilton vector and in particular, see the references \cite{prec1,ham6,ham7} for its use in the calculation of the perihelion  precession angle).

The Hamilton vector, $\vec{u}$, is another conserved vector of the Kepler problem; it not independent from the more known Runge-Lenz $\vec{R}$ vector, in fact they are connected by the relation \cite{prec1}:\footnote{In this equation it is evident that the Hamilton vector is a conserved quantity, since it is linked to other constants of the motion.}
\begin{equation*}
\vec{R} = \vec{u} \, \times \vec{L}
\end{equation*}
where $\vec{L}$ is the angular momentum.

Given a particle of mass $m$ subjected to the gravitational force of a central object of mass $M$, the Hamilton vector and its magnitude are given respectively by \cite{prec1}:
\begin{subequations}
\begin{align}
\vec{u} &= \dot{\vec{x}} - \dfrac{GM}{h} \, \vec{e}_{\varphi}\\  \nonumber
&= \dot{r} \, \vec{e}_r + \left( r\, \dot{\varphi} -  \dfrac{GM}{h} \right) \,  \vec{e}_{\varphi}
\end{align}
\begin{equation}
|u| = \dfrac{GM}{h} \, e
\end{equation}
\end{subequations}
where $h$ is the magnitude of the angular momentum per unit mass and $e$ is the eccentricity of the orbit and $\vec{e}_r$ and $\vec{e}_{\varphi}$ are unit vectors ($\varphi$  is the polar coordinate and $r$ is the radial one).

The precession rate for the Hamilton vector is given by \cite{prec1,merc}:
\begin{align*}
\vec{w}&= \dfrac{\vec{u} \, \times \, \dot{\vec{u}}}{|u|^2} \\
\intertext{where, (using the first order equation of motion):}
\dot{\vec{u}} &= - \dfrac{\dot{\vec{x}}}{c} \,\dfrac{d^2}{dt^2}{}^3\widetilde{\mathcal{K}}_{(1)} = - \, \dfrac{1}{c} \, \Big( \dot{r} \, \vec{e}_r + r \, \dot{\varphi} \, \vec{e}_{\varphi} \Big) \, \dfrac{d^2}{dt^2} \, {}^3\widetilde{\mathcal{K}}_{(1)}
\end{align*}
so, we find:
\begin{equation}\label{eq:rate}
\vec{w} = - \, \left( \dfrac{h}{G M e} \,  \dfrac{\dot{r}}{c} \, \dfrac{d^2}{dt^2} \, {}^3\widetilde{\mathcal{K}}_{(1)}\right) \; \vec{k}.
\end{equation}
where $\vec{k}$ is the unit vector orthogonal to the plane of the orbit.

The precession is the integral of the rate \eqref{eq:rate} over a period $T$:
\begin{equation}
\Delta \, \theta_P = \int_0^T {|w| \,dt} = - \dfrac{1}{c} \, \dfrac{h}{GMe} \, \int_0^T{\dot{r} \, \dfrac{d^2}{dt^2} \, {}^3\widetilde{\mathcal{K}}_{(1)} \, dt}
\end{equation}
where  $\dot{r}$ is the first order radial velocity (see \cite{gold}). The last equation can be reformulated as an integral over the radial coordinate $r$:
\begin{equation}\label{eq:preces1}
\Delta \, \theta_P = -\dfrac{1}{c} \, \dfrac{h}{GMe} \, \int_{r_{-}}^{r_{+}}{\dfrac{d^2}{dt^2} \, {}^3\widetilde{\mathcal{K}}_{(1)} \, dr}
\end{equation}

Now we can introduce the variable $z$ \cite{prec,prec1,gold}:
\begin{equation*}
\dfrac{S}{r} =  1 + e\,\cos\varphi  = 1+ez
\end{equation*}
where $S= h^2/GM$ is the \emph{semilatus rectum} \cite{prec,prec1}. With this variable, we have:
\begin{equation}\label{eq:precessione}
\Delta \theta_P = \dfrac{2}{c} \, \dfrac{h}{GM} \, \int_{-1}^{1}{ \dfrac{S}{(1+ez)^2} \,  \left( \dfrac{d^2}{dt^2}{}^3\widetilde{\mathcal{K}}_{(1)} \right) dz}
\end{equation}
where:
\begin{align} \nonumber
\dfrac{d^2}{dt^2} \, {}^3\widetilde{\mathcal{K}}_{(1)} &= \left( {\partial_t}^2 + 2 \, \dot{r} \, \partial_r + \dot{r}^2 \, {\partial_r}^2 \right) \, {}^3\widetilde{\mathcal{K}}_{(1)}=\\ \label{eq:der}
&= \left[ {\partial_t}^2 + 2 \, \dot{r} \, \left( \dfrac{\partial z}{\partial r}\right) \, \partial_z + \dot{r}^2 \, \left(\dfrac{\partial z}{\partial r} \right)^2 \, {\partial_z}^2 \right] \, {}^3\widetilde{\mathcal{K}}_{(1)} = \\ \nonumber
&= \left[ {\partial_t}^2 - 2 \, \dot{r} \, \dfrac{(1+e z)^2}{e \,S} \, \partial_z + \dot{r}^2 \, \left( \dfrac{(1+e z)^2}{e \,S} \right)^2 \, {\partial_z}^2\right] \, {}^3\widetilde{\mathcal{K}}_{(1)}
\end{align}
Finally, from \cite{gold}, we find $\dot{r} = h \, \sqrt{1 - z^2}$.

\subsection{Results}

We use the data reported in table \ref{tab:pre_pla} taken from \cite{sol, sol2}: the error in Jupiter measurement is orders of magnitude bigger than the others and therefore it shall not be considered.

\begin{table}[t]
\centering
\begin{tabular}{c|cc}
Planet & Precession & Error \\
 & ($\times 10^{-11}$ rad) & ($\times 10^{-11}$ rad) \\
\hline
Mercury & -1.5 & 3.5 \\
Venus & 7.8 & 4.8 \\
Earth & 0.9 & 0.9 \\
Mars & -0.19 & 0.38 \\
Jupiter & 3400 & 1632 \\
Saturn & -45 & 67
\end{tabular}
\caption{Precessions of planets per orbital period and their errors (see \cite{sol,sol2}).}\label{tab:pre_pla}
\end{table}

\begin{figure}[ht]
\centering
\includegraphics[width=0.45\textwidth,keepaspectratio]{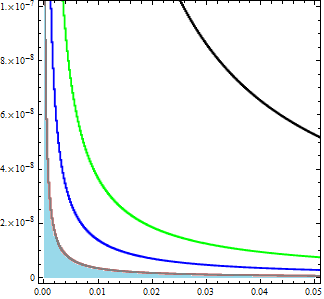}
\caption{Constraints on parameters imposed by planets' precessions: on the abscissa $\delta$ on the ordinate $\mu$; black line is Mercury, green Venus, blue Earth; Mars and Saturn constraints are almost overlapping (brown line). The allowed parameter space is colored in light blue.}\label{fig:sol_sys}
\end{figure}

Our results are given in figure \ref{fig:sol_sys}: the black line is the upper limit set by Mercury, the green one by Venus, the  blue one by the Earth; Mars and Saturn are almost overlapping (brown line);  the allowed parameter space is colored in light blue.

\section{Combining both methods}
\label{sec:both}
We can now combine the constriaints found in the previous sections.

\begin{figure}[htb]
\centering
\includegraphics[scale=0.9]{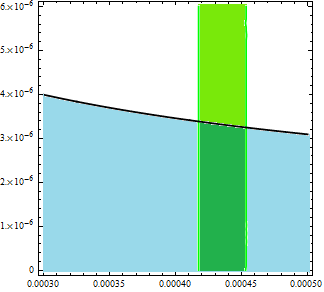}
\caption{Combination of constraints: ligth green stripe is the constraint from the S2 star, light blue area is the allowed parameter space from the Solar System, darker green is the combined constraint. The red dot is the best fit for the S2 orbit.}\label{fig:both}
\end{figure}

Referring to figure \ref{fig:both}: the black line is the combined upper limit imposed by Saturn and Mars, the light blue area is the same as in figure \ref{fig:sol_sys}, the light green stripe is the area allowed by the orbit fit, finally, the darker green area is the combined constraint. In formulae we have:
\begin{equation}\label{eq:fin_constr}
\left\{\begin{array}{l}
4.2 \,\times \,  10^{-4} \, \text{AU}\,\lesssim \, \delta \, \lesssim \, 4.6 \, \times \, 10^{-4} \, \text{AU} \\
\\
\mu \, \lesssim \, 3.5 \, \times \, 10^{-6} \, \text{AU}^{-1}.
\end{array}\right.
\end{equation}

\section{Conclusion}

ADM tetrad gravity is an Hamiltonian reformulation of General Relativity that gives new insight on Dark Matter problem: Dark Matter can be seen as an inertial effect due to the choice of the 3+1 splitting of the spacetime. 

In this work we suggested a possible ansatz for the first order of the trace of the extrinsic curvature: a Yukawa-like \emph{force} with a linear time dependence. In this ansatz there are two free parameters: the \emph{strength} of the Yukawa coupling $\delta$ and its length scale. We then imposed constraints on the values of these parameters.

In order to constraint the parameter space, we fitted the orbit of the S2 star around the black hole in the galactic center to the one given by ADM tg equations of motion and confronted the result with a pure Keplerian orbit (see also \cite{capo}).

We find that the ADM tg fit is only marginally better than that given by the simple Newtonian potential, in fact the reduced $\chi^2$ are 1.5477 for the Keplerian orbit  and 1.516890 for the ADM tg with two additional degrees of freedom. 

We then used the perihelia of some of the planets of the Solar System to impose upper limits on $\delta$ and $\mu$.

Both the previous methods give unsatisfactory results, but their combination  sets very tight limits on the \emph{strength} of the Yukawa coupling, $4.2 \,\times \,  10^{-4} \, \text{AU}\,\lesssim \, \delta \, \lesssim \, 4.6 \, \times \, 10^{-4} \, \text{AU}$, and an upper limit for the (inverse) scale length, $\mu \, \lesssim \, 3.5 \, \times \, 10^{-6} \, \text{AU}^{-1}$.


\begin{thebibliography}{100}
\bibitem{adm1} L. Lusanna and S. Russo, Gen. Rel. Grav., \textbf{34}, 189 (2002);
\bibitem{adm2} R. De Pietri, L. Lusanna, L. Martucci, S. Russo, Gen. Rel. Grav., \textbf{34}, 877 (2002);
\bibitem{adm3} D. Alba, L. Lusanna, Gen. Rel. Grav. \textbf{39}, 2149 (2007);
\bibitem{adm4} D. Alba, L. Lusanna, Can. J. of Phys. \textbf{90}, 1017 (2012)
\bibitem{adm5}  D. Alba, L. Lusanna, Can. J. of Phys. \textbf{90}, 1077 (2012);
\bibitem{adm6}  D. Alba, L. Lusanna, Can. J. of Phys. \textbf{90}, 1131 (2012);
\bibitem{adm7} L. Lusanna, presented at the Black objects in Supergravity - Springer Proceedings in Physics, ed. by S. Bellucci, p. 267;
\bibitem{adm8} S. Deser, C. Isham, Phys. Rev. D \textbf{14}, 2505 (1976);
\bibitem{adm} Arnowitt, R., Deser, S., Misner, C. \emph{The dynamics of General Relativity}, in \emph{Gravitation: an introduction to current research}, ed. L. Witten (Wiley, New York, 1962); republished in Gen. Rel. Grav., \textbf{40}, 1997-2027 (2008);
\bibitem{capo2} S. Capozziello, M. de Laurentis, Annalen der Physik \textbf{524}, 545 (2012)
\bibitem{gill1} Gillessen, S. \emph{et al.}, ApJ. \textbf{692}: 1075 - 1109 (2009);
\bibitem{gill2} Gillessen, S. \emph{et al.}, ApJL. \textbf{707}: L114 - L117 (2009);
\bibitem{cpp} Press H. W., Teukolsky, S.A., Vetterling, W.T., Flannery, B.P., \emph{Numerical recipes in C - The art of scientific computing}, 2nd edition (1992), Cambridge University Press;
\bibitem{capo} Capozziello, S. \emph{et al.}, Phys. Rev. D, \textbf{90} 044052 (2014);
\bibitem{sand} Sanders, R. H., A. \& A., \textbf{136}, L21-L23 (1984);
\bibitem{prec} Adkins, G. S. and McDonnel, J., Phys. Rev D, \textbf{75}, 082001 (2007);
\bibitem{prec1} Chashchina, O. I. and Siligadze, Z. K., Phys. Rev. D \textbf{77}, 107502 (2008);
\bibitem{ham1} Mart\'inez-y-Romero, R. P., N\'u\~{n}ez-Y\'epez, H. N. and Salas-Brito, A. L., Eur. J. Phys, \textbf{14}, 71 (1993);
\bibitem{ham2}  R. P., N\'u\~{n}ez-Y\'epez,H.N. and Salas-Brito, A.L., Eur. J. Phys, \textbf{21} L39 (2000);
\bibitem{ham3} Munoz, G., Am. J. Phys, \textbf{71}, 1292 (2003);
\bibitem{ham4} Wheeler, J. T., Can. J. of Phys. \emph{83}, 91 (2005);
\bibitem{ham5} H. Goldstein, Am. J. Phys., \textbf{44}, 1123 (1976);
\bibitem{ham6} B. Davies, Am. J. Phys., \textbf{51}, 909 (1983);
\bibitem{ham7} Ebner, D., Am. J. Phys., \textbf{53}, 374 (1985);
\bibitem{merc} Stewart, M. G., Am. J. Phys., \textbf{73}, 730 (2005);
\bibitem{gold} Goldstein, H., Poole, S. and Safko, J. \emph{Classical Mechanics}, 3rd edition, Addison Wesley (2001);
\bibitem{sol} E. V. Pitjeva, N. P. Pitjev, MNRAS, \textbf{432}, 3431-3437 (2013);
\bibitem{sol2} E. V. Pijeva, Astronomy Letters, \textbf{39}, 141-149 (2013);
\bibitem{sol3} I. B. Khriplovich and E. V. Pitjeva, IJMP D, \textbf{15}, 615-618 (2006);
\bibitem{sol4} I. B. Khriplovich, IJMP D , \textbf{16}, 1475-1478 (2007).
\end{thebibliography}
\end{document}